# Interface controlled antiferromagnetic tunnel junctions

Liu Yang,[1,2,*] Yuan-Yuan Jiang,[1,2,*] Xiao-Yan Guo,[1,†] Shu-Hui Zhang,[3] Rui-Chun Xiao,[4] Wen-Jian Lu,[1] Lan Wang,[7] Yu-Ping Sun,[5,1,6] Evgeny Y. Tsymbal,[8,‡] and Ding-Fu Shao[1,§]

[1] *Key Laboratory of Materials Physics, Institute of Solid State Physics, HFIPS, Chinese Academy of Sciences, Hefei 230031, China*
[2] *University of Science and Technology of China, Hefei 230026, China*
[3] *College of Mathematics and Physics, Beijing University of Chemical Technology, Beijing 100029, China*
[4] *Institute of Physical Science and Information Technology, Anhui University, Hefei 230601, China*
[5] *Anhui Province Key Laboratory of Low-Energy Quantum Materials and Devices, High Magnetic Field Laboratory, HFIPS, Chinese Academy of Sciences, Hefei, Anhui 230031, China*
[6] *Collaborative Innovation Center of Microstructures, Nanjing University, Nanjing 210093, China*
[7] *Department of Physics, School of Physics, Hefei University of Technology, Hefei, Anhui, 230009, China*
[8] *Department of Physics and Astronomy & Nebraska Center for Materials and Nanoscience, University of Nebraska, Lincoln, Nebraska 68588-0299, USA*

[*] These authors contributed equally to this work.
[†] xyguo@issp.ac.cn; [‡] tsymbal@unl.edu; [§] dfshao@issp.ac.cn

Magnetic tunnel junctions (MTJs) are the key building blocks of high-performance spintronic devices. While conventional MTJs rely on ferromagnetic (FM) materials, employing antiferromagnetic (AFM) compounds can significantly increase operation speed and packing density. Current prototypes of AFM tunnel junctions (AFMTJs) exploit antiferromagnets either as spin-filter insulating barriers or as metal electrodes supporting bulk spin-dependent currents. Here, we highlight a largely overlooked AFMTJ prototype, where bulk-spin-degenerate electrodes with an A-type AFM stacking form magnetically uncompensated interfaces, enabling spin-polarized tunneling currents and a sizable tunneling magnetoresistance (TMR) effect. Using first-principles quantum-transport calculations and the van der Waals (vdW) metal $Fe_4GeTe_2$ as a representative A-type AFM electrode, we demonstrate a large negative TMR arising solely from the alignment of interfacial magnetic moments. This prototype of AFMTJs can also be realized with various non-vdW A-type AFM metals that support roughness-insensitive surface magnetization. Beyond TMR, AFMTJs based on A-type antiferromagnets allow convenient switching of the Néel vector, opening a new paradigm for AFM spintronics that leverages spin-dependent properties at AFM interfaces.

Spintronics exploits spin-dependent transport properties for information processing and storage [1]. A key device in this field is the magnetic tunnel junction (MTJ), typically composed of two metallic ferromagnetic (FM) electrodes separated by a nonmagnetic (NM) insulating barrier [2-4]. The tunneling conductance depends on the relative magnetization orientation of the electrodes, giving rise to the tunneling magnetoresistance (TMR) effect [5-7]. Additionally, the tunneling spin-polarized currents can exert spin-transfer torque (STT) on FM magnetization, enabling its rotation and switching [8-10]. A different approach employs spin-filter MTJs, where an insulating FM barrier layer controls spin-filter tunneling [11-16]. Substantial efforts have been paid to improve density and operation speed of spintronic devices based on conventional MTJs. However, these MTJs suffer from the limitations associated with the stray fields and relatively slow (GHz frequency range) spin dynamics intrinsic to ferromagnets.

These limitations of ferromagnets can be addressed by using antiferromagnetic (AFM) tunnel junctions (AFMTJs) [17], as antiferromagnets exhibit ultrafast spin dynamics in the THz frequency range and produce no stray fields [18-20]. However, due to their lack of net magnetization, antiferromagnets have long been considered inefficient as functional elements in MTJs. Nonetheless, over the past decade, two prototypes of AFMTJs demonstrating sizable TMR have been developed. Prototype-I utilizes a thin layer of van der Waals (vdW) AFM insulator as a tunnel barrier (Fig. 1(A)). This a spin-filter AFMTJ employs an applied magnetic field to switch the AFM ordering of the barrier material into a metastable FM state, resulting in a giant spin-filter TMR effect [21-28]. Although this FM state is volatile, it can be made non-volatile if the AFM layer has an in-plane Néel vector and is twisted [29]. Prototype-II relies on spin-dependent current originating from the AFM bulk electronic structure (Fig. 1(B)) [30-44]. The TMR effect in these AFMTJs is associated with the momentum-dependent spin polarization of collinear spin-split antiferromagnets [45-55], now known as altermagnets [56,57], and noncollinear antiferromagnets [30-34]. It can also arise from sublattice-dependent spin polarization determined by AFM stacking in real space [58,59]. The specific growth direction of the AFM electrodes is critical to supporting the spin-dependent current. Recently, room-temperature TMR has been demonstrated in AFMTJs with noncollinear AFM electrodes [35,36].

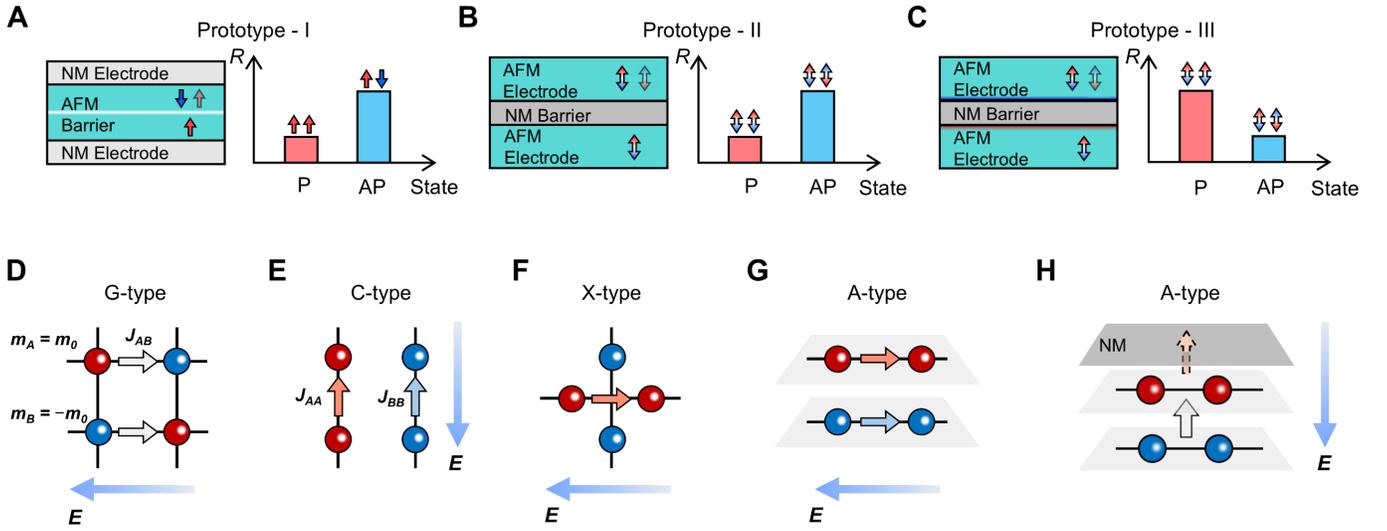

**Figure. 1: AFMTJ prototypes.** (A) Prototype-I AFMTJs with NM electrodes and AFM barrier, where magnetic phase transition in the barrier controls the resistance states. (B) Prototype-II AFMTJs with AFM electrodes and NM barrier, where a match (mismatch) of the AFM bulk states results in low (high) resistance for the P (AP) state of the AFMTJ. (C) Prototype-III AFMTJs with AFM electrodes and NM barrier, where a match (mismatch) of the AFM interfacial states results in low (high) resistance for the AP (P) state. (D) G-type AFM stacking, where the net current is primarily determined by the spin-neutral inter-sublattice currents. (E) C-type AFM stacking, where the global current along the FM chain direction corresponds to staggered Néel spin currents. (F) X-type AFM stacking, where the global current along one of the chain directions is determined by an isolated Néel spin current. (G) A-type AFM stacking, where an in-plane net current represents staggered Néel spin currents. The spin-dependent currents in panels (E-G) can be used in prototype-II AFMTJs. (H) A-type AFM stacking, where a magnetically uncompensated interface drives the spin-dependent tunneling current and generates TMR in prototype-III AFMTJs. In panels (D-H), the solid lines connect the nearest neighbor atoms, the red and blue balls denote the opposite magnetic sublattices $m_A$ and $m_B$, respectively, the hollow arrows denote the local currents between nearest neighbor atoms, and the solid blue arrows denote the electric fields. Here the gray color indicates the vanishing spin polarization of the inter-sublattice current $J_{AB}$, and the red and blue colors indicate the intra-sublatice currents $J_{AA}$ and $J_{BB}$ of opposite spin polarization, respectively.

Although these AFMTJ prototypes are promising, they exclude a wide range of antiferromagnets that neither exhibit magnetic phase transitions nor support spin-dependent currents in their bulk. To address this limitation, we emphasize a previously discussed but largely overlooked AFMTJ prototype, where A-type AFM stacking of the electrodes creates magnetically uncompensated interfaces capable of generating spin-dependent tunneling currents and producing a sizable TMR effect (Fig. 1(C)). Based on first-principles quantum-transport calculations and the two-dimensional (2D) vdW antiferromagnet $Fe_4GeTe_2$ as a representative A-type electrode, we demonstrate a large TMR solely attributed to the interface-driven tunneling spin current. Since A-type AFM stacking can potentially be realized in most antiferromagnets by selecting appropriate growth orientations, we argue that this AFMTJ prototype can be created using a broad range of AFM metals with roughness-robust interfaces.

**Stacking-dependent spin currents in antiferromagnets**

Electric currents in collinear antiferromagnets can be intuitively described in terms of charge currents flowing within a given magnetic sublattice, i.e., the intra-sublattice currents, and those flowing between the two magnetic sublattices, i.e., the inter-sublattice currents [58]. The intra-sublattice currents are expected to be spin-polarized, while the inter-sublattice currents to be neutral. For example, in antiferromagnets with a G-type stacking, where the magnetic moments are aligned antiparallel between all nearest neighbors, the intra-sublattice currents are negligible due to the large hopping distance between the atoms that belong to the same sublattice. As a result, the net current is spin independent due to being contributed mainly by the inter-sublattice currents (Fig. 1(D)). This is different from antiferromagnets with C-, X-, and A-type stackings, where the building blocks are FM chains or FM layers (Fig. 1(E-G)) [59,60]. For the C- (A-) type stacking, an electric field parallel to the chains (layers) generates a globally spin-neutral current, representing a staggered Néel spin current flowing within the chains (layers), since the inter-sublattice currents are negligible (Fig. 1(E, G)) [58]. For the X-type stacking proposed recently, where FM chains of the two magnetic sublattices form a cross pattern, an electric field along one chain generates a net current which is majorly composed of an isolated Néel spin current on a single sublattice (Fig. 1(F)) [59]. The X-type stacking offers an intuitive picture for spin currents generated by altermagnetic [56,57] and non-altermagnetic [61] spin splitting [59].

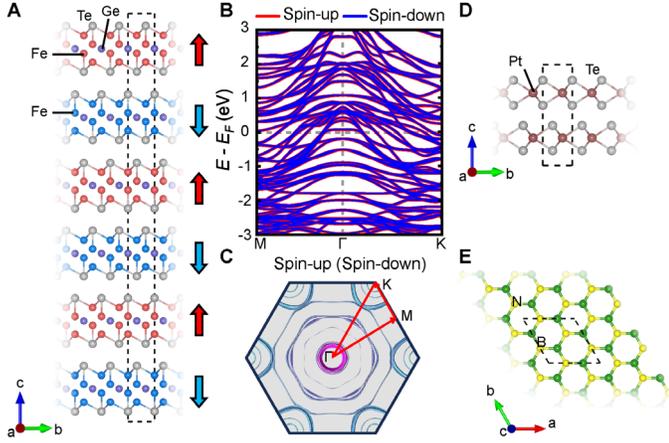

**Figure. 2: 2D vdW materials for prototype-III AFMTJs. (A)** Side view of the 2D vdW A-type AFM metal $Fe_4GeTe_2$. Fe atoms are indicated by red (blue) balls have positive (negative) magnetic moments. **(B, C)** The spin-resolved band structure **(B)** and top view of the Fermi surface **(C)** of bulk $Fe_4GeTe_2$. **(D, E)** Atomic structure of 2D vdW metal 2H-$PtTe_2$ **(D)** and insulator BN **(E)**. Pt, Te, B, and N atoms are indicated by brown and gray, yellow, and green balls, respectively.

This simple analysis suggests that the spin-dependent current, regardless of its net spin polarization, can be supported by an antiferromagnet if the intra-sublattice currents are much stronger than the inter-sublattice currents in the transport direction. C-, A-, and X-type antiferromagnets are thus suitable as electrodes in AFMTJs of prototype-II (Fig. 1(B)), where the transport spin polarization is generated in bulk of AFM metals [17,58,59].

In addition to the bulk-induced spin polarization of AFM electrodes, the spin-dependent tunneling current in AFMTJs can be produced by their interfaces. For example, in a bulk A-type antiferromagnet, an out-of-plane electric field can only generate spin-neutral inter-sublattice currents (Fig. 1(H)). However, when using such antiferromagnet as electrodes in a tunnel junction, the tunneling current is expected to be spin polarized due to the uncompensated magnetic moment of the interfacial layer. This property can be functionalized in AFMTJs of prototype-III (Fig. 1(C)). A notable difference between this and other AFMTJ prototypes is the TMR behavior. In prototypes-I and -II, an AFMTJ with the parallel- (P-) aligned Néel vector always has higher conductance ($G_P$) than that with the antiparallel- (AP-) aligned Néel vector ($G_{AP}$), resulting in a positive TMR ratio defined by $TMR = \frac{G_P - G_{AP}}{\min\{G_P, G_{AP}\}}$. Fig. 1(A, B)). On the contrary, in a prototype-III AFMTJ, it is the interface termination that determines the TMR ratio, rather than bulk electrodes. While a positive TMR is expected if the P- (AP-) aligned Néel vector in the electrodes is associated with P- (AP-) aligned interfacial magnetic moments, a negative TMR will occur if the P- (AP-) aligned Néel vector corresponds to AP- (P-) aligned interfacial moments (Fig. 1(C)).

While prototype-III AFMTJs were proposed some time ago [62-64], their potential has largely been overlooked. This issue arose due to interface roughness, which was considered unavoidable and detrimental to interfacial spin polarization and the associated TMR. However, this problem may be solved by employing A-type antiferromagnets with roughness-insensitive surface magnetization coupled to the Néel vector [65-67]. The most straightforward solution is to use A-type AFM electrodes composed of 2D vdW materials, where vdW gaps ensure flat surface terminations. These AFMTJs can be fabricated using vdW-assembly techniques.

**Interface effect on spin-dependent tunneling current**

We first demonstrate the interface-driven spin-dependent transport of a representative vdW A-type AFM metal based on first-principles calculations. Figure 2(A) shows the structure of vdW metal $Fe_4GeTe_2$, where each monolayer is centrosymmetric and has FM-aligned Fe moments. While $Fe_4GeTe_2$ is an FM material, it is known that the A-type AFM ordering can be induced by Co doping [68]. In the following, for simplicity, we consider pristine $Fe_4GeTe_2$ with A-type AFM stacking as a representative electrode to demonstrate AFMTJs of prototype-III. This simplification does not influence our conclusions because, as shown in Supplemental Note S4, the effect of Co largely produces electron doping, and therefore the results obtained for the AFM-ordered $Fe_4GeTe_2$ can be extrapolated to those for Co-doped $Fe_4GeTe_2$ by shifting the Fermi energy to higher energies.

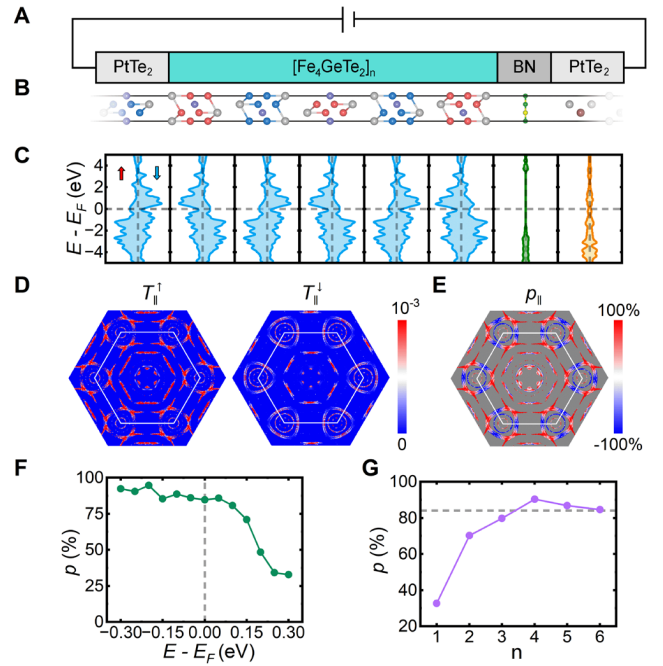

**Figure. 3: Interfacial effect on spin-dependent current. (A)** Schematic of the $PtTe_2/[Fe_4GeTe_2]_n/BN/PtTe_2$ tunnel junction. **(B, C)** Atomic structure **(B)** and monolayer-resolved density of states **(C)** of the tunnel junction. **(D-F)** Spin- and $\vec{k}_\parallel$-dependent transmissions $T_\parallel^\uparrow(\vec{k}_\parallel)$ and $T_\parallel^\downarrow(\vec{k}_\parallel)$ **(D)**, $\vec{k}_\parallel$-dependent spin polarization $p_\parallel(\vec{k}_\parallel)$ at fermi energy $E_F$ **(E)**, and total transport spin polarization $p$ as a function of energy **(F)** in $Fe_4GeTe_2/BN/PtTe_2$ tunnel junction. **(G)** Total transport spin polarization $p$ at $E_F$ as a function of different layers $n$ in $PtTe_2/[Fe_4GeTe_2]_n/BN/PtTe_2$ tunnel junction. These results suggest the giant and robust spin polarizations of the interface-driven tunneling currents.

The A-type AFM stacking of bulk $Fe_4GeTe_2$ is characterized by coexistence of inversion symmetry $\hat{P}$ and combined symmetry $\hat{P}\hat{T}\hat{t}$ in bulk material, where $\hat{T}$ is time reversal symmetry and $\hat{t}$ is translation symmetry. A thin film of $Fe_4GeTe_2$ with even number of monolayers, however, is non-centrosymmetric and hosts combined $\hat{P}\hat{T}$ symmetry. These symmetries enforce spin-degenerate electronic structure of both bulk (Fig. 2(B)) and even-monolayer film of $Fe_4GeTe_2$. Despite this spin degeneracy, when an electric field is applied in the plane, staggered Néel spin currents are generated along the parallel-aligned FM-ordered monolayers of $Fe_4GeTe_2$ [58]. The situation is different when an electric field is applied out of the plane. As seen from Figure 2(C), the Fermi surface of bulk $Fe_4GeTe_2$ is spin degenerate along the out-of-plane direction. At the same time, due to alternating magnetic moments along the $c$ axis (Fig. 2(A)), there are no Néel spin currents along the out-of-plane direction, indicating spin-independent transport. As a result, no bulk contribution to the spin dependence of the tunneling current is thus expected when using $Fe_4GeTe_2$ as the electrodes in AFMTJs.

To reveal the interface effect on spin-polarized tunneling, we construct a tunnel junction $PtTe_2/[Fe_4GeTe_2]_n/BN/PtTe_2$ (Fig. 3(A)), where NM vdW metal 2H-$PtTe_2$ (Fig. 2(D)) [69] is used as leads, NM insulating monolayer BN (Fig. 2(E)) [70] is used as barrier, and $n$-monolayer $Fe_4GeTe_2$ is used as a spacer layer. The Néel vector of $Fe_4GeTe_2$ are set parallel to the magnetic moments of the interfacial monolayer (Fig. 3(B)). Figure 3(C) shows the monolayer-resolved density of states of this heterostructure. We find that a wide band gap of BN is well maintained, and the Fermi energy $E_F$ is located deeply inside the band gap.

We first calculate transmission of the tunnel junction with an infinite $n$ which corresponds to a $Fe_4GeTe_2/BN/PtTe_2$ junction where the left electrode is bulk $Fe_4GeTe_2$. Despite the spin-degenerate Fermi surface in bulk $Fe_4GeTe_2$ (Fig. 2(C)), we find that the $\vec{k}_\parallel$-dependent transmissions $T_\parallel^\sigma(\vec{k}_\parallel)$ are different for the up-spin ($\sigma = \uparrow$) and down-spin ($\sigma = \downarrow$) channels (Fig. 3(C)), leading to a finite $\vec{k}_\parallel$-dependent spin polarization $p_\parallel(\vec{k}_\parallel) = \frac{T_\parallel^\uparrow - T_\parallel^\downarrow}{T_\parallel^\uparrow + T_\parallel^\downarrow}$ (Fig. 3(D)). Since bulk electronic structure of $Fe_4GeTe_2$ is spin-degenerate, the predicted spin polarization is solely driven by the uncompensated magnetic moments at the interface of $Fe_4GeTe_2$ with BN. Overall, we find that $T_\parallel^\uparrow(\vec{k}_\parallel) > T_\parallel^\downarrow(\vec{k}_\parallel)$ for most $\vec{k}_\parallel$, resulting in the total transmission being much larger for the up-spin channel ($T^\uparrow$) than for the down-spin channel ($T^\downarrow$) and the net spin polarization $p = \frac{T^\uparrow - T^\downarrow}{T^\uparrow + T^\downarrow} = 84\%$ at $E_F$. As is evident from Figure 3(F), $p$ is maintained to be large and positive within a wide energy window around $E_F$.

We then calculate spin-dependent transmission of the $PtTe_2/[Fe_4GeTe_2]_n/BN/PtTe_2$ tunnel junction with a finite $n$. As seen from Figure 3(G), the $Fe_4GeTe_2$ spacer layer in the 2D limit ($n = 1$) produces a small $p$, due to the interfacial monolayer being proximitized with $PtTe_2$. For $n \geq 2$, $p$ increases and converges to the value of bulk $Fe_4GeTe_2$ when $n = 6$. We find that $p$ at $E_F$ is always positive regardless of $n$ being odd or even. These facts demonstrate the interface origin of the spin-polarized current.

**TMR in AFMTJs with $Fe_4GeTe_2$ electrodes**
The sizable interface effect on spin-dependent tunneling current indicates that $Fe_4GeTe_2$ can be used in electrodes of prototype-

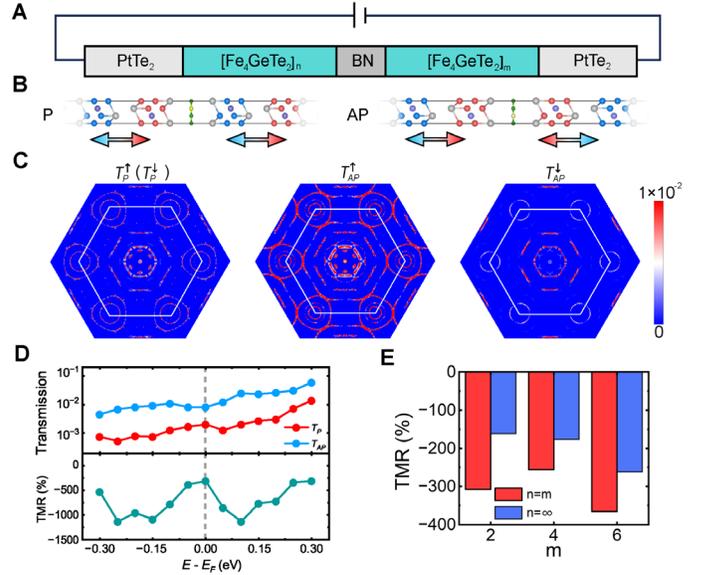

**Figure. 4: TMR of $PtTe_2/(Fe_4GeTe_2)_n/BN/(Fe_4GeTe_2)_m/PtTe_2$ AFMTJ. (A, B)** Schematics of the AFMTJ **(A)** and its structure for P and AP states **(B)**. **(C)** Calculated $\vec{k}_\parallel$-resolved transmission ($T$) at $E_F$ for P and AP states and for the case of $n = m = 2$. **(D)** Total transmission for P and AP states (top panel) and TMR ratio (bottom panel) as functions of energy for $n = m = 2$, indicating the large TMR sustained within a broad energy window around $E_F$. **(E)** TMR ratio for $n = m$, which remains sizable for different thicknesses of electrodes.

III AFMTJs. We thus construct PtTe$_2$/(Fe$_4$GeTe$_2$)$_n$/BN/(Fe$_4$GeTe$_2$)$_m$/PtTe$_2$ AFMTJ as shown in Figures 4(A, B). We first calculate transmission of the AFMTJ with the same Fe$_4$GeTe$_2$ spacer layer thickness on the two sides ($n = m$). Figure 4(C) shows the $\vec{k}_\parallel$-dependent transmissions, $T_P^\sigma(\vec{k}_\parallel)$ and $T_{AP}^\sigma(\vec{k}_\parallel)$, for the P and AP states, respectively, for AFMTJ with $n = m = 2$. It is seen that $T_P^\uparrow(\vec{k}_\parallel) = T_P^\downarrow(\vec{k}_\parallel)$, which is due to the $\hat{P}\hat{T}$ symmetry being preserved in the AFMTJ with the P-aligned Néel vector (Fig. 4(B)). The interfacial magnetic moments are AP-aligned in this case, suppressing the spin-dependent tunneling current. At the same time, we find that $T_{AP}^\uparrow(\vec{k}_\parallel)$ is enhanced and $T_{AP}^\downarrow(\vec{k}_\parallel)$ is reduced for the AP state, due to the interfacial moments being P-aligned (Fig. 4(C)). These results indicate that the alignment of the interfacial moments controls the spin-dependent tunneling current in this AFMTJ. As a result, the total transmission for the P state ($T_P$) appears to be much smaller than that for the AP state ($T_P$) (Fig. 4(D)). The predicted TMR ratio is $TMR = \frac{T_P - T_{AP}}{\min\{T_P, T_{AP}\}} = -300\%$ at $E_F$. The large negative TMR is sustained within a broad energy window around $E_F$ (Fig. 4(D)).

For Fe$_4$GeTe$_2$ spacer layer thickness of $n = m = 4$ and 6, we obtain TMR ratios of about $-250\%$ and $-360\%$, respectively (Fig. 4(E)). For bulk Fe$_4$GeTe$_2$ representing the left electrode ($n = \infty$) and a thin Fe$_4$GeTe$_2$ spacer layer in the right electrode ($m$ is small), the TMR ratios are predicted to be between $-150\%$ and $-300\%$ (Fig. 4(E)). These results indicate the minor influence of Fe$_4$GeTe$_2$ layer thickness on TMR, due to the dominant contribution of the interfacial magnetic moments rather than bulk effects. Moreover, we find the TMR ratio can be further enhanced by increasing the thickness of BN barrier (see Supplemental Note S2).

**Discussion**

Previous theoretical calculations of spin transport in AFMTJs [62-64,71] have predicted either a very small zero-bias TMR [62-64] or TMR oscillating in sign as a function of energy [71]. These results indicate competing bulk- and interface-induced contributions to the spin polarization of the tunneling current. In some cases, TMR is dominated by a spin-neutral contribution from the bulk electrodes, leading to small TMR, while in others, the spin-dependent bulk contribution opposes the interface-driven spin polarization, causing TMR to oscillate in sign with energy. This behavior is illustrated in our modeling of a hypothetical Cr/Si/Cr AFMTJ (see Supplemental Note S8), where TMR oscillates in sign as a function of energy but can be made consistently positive or negative across all energies by applying strain that enhances or reduces the bulk-controlled spin-dependent current, respectively. In contrast, the prototype-III AFMTJs designed in our work exhibit a large TMR solely

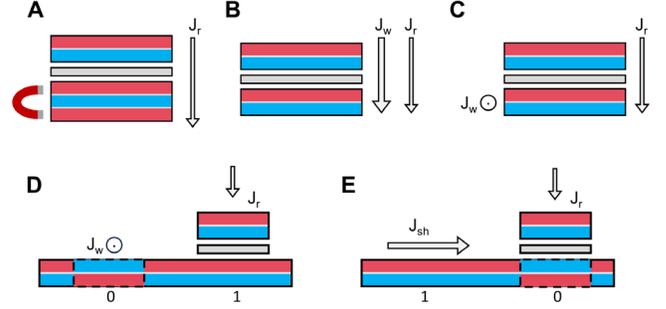

**Figure. 5: Write-in and read-out of the prototype-III of AFMTJs.** The gray stripe denotes the NM barrier. The red and blue stripes denote the layers in A-type AFM electrodes with positive and negative moments, respectively. **(A)** The switching of the bottom odd-layer electrode by a magnetic field. **(B)** The switching of the bottom electrode by a spin transfer torque via an out-of-plane writing current $J_w$. **(C)** The switching of the bottom electrode by the Néel torques via an in-plane $J_w$. **(D, E)** An AFM racetrack memory made of an A-type antiferromagnet, where the AFM domains are created by an in-plane $J_w$ **(D)** and transported by an in-plane shifting current $J_{sh}$ **(E)**. An AFMTJ is implemented for detecting the domain via an out-of-plane reading current $J_r$. The 0 and 1 represent low- and high-resistance states, respectively. The arrows denote the currents, where the widths of arrows qualitatively represent the associated current densities.

controlled by interface magnetic moments, showing no sign change as a function of energy.

The robustness of the predicted TMR against the position of the Fermi energy implies similar behavior in realistic AFMTJs based on Fe$_4$GeTe$_2$ electrodes doped with Co, which is necessary to induce antiferromagnetism in Fe$_4$GeTe$_2$. Since Co doping adds valence electrons and raises the Fermi energy, we calculated the transport properties of PtTe$_2$/[(Fe$_{0.67}$Co$_{0.33}$)$_4$GeTe$_2$]$_n$/BN/[(Fe$_{0.67}$Co$_{0.33}$)$_4$GeTe$_2$]$_m$/PtTe$_2$ AFMTJs, explicitly modeling Co doping by the virtual crystal approximation. The results closely align with those obtained for Fe$_4$GeTe$_2$-based AFMTJs, as discussed in Supplemental Note S4. Moreover, prototype-III AFMTJs can be realized using A-type AFM electrodes beyond Fe$_4$GeTe$_2$. A-type AFM phases with high Néel temperatures have been reported in 2D vdW AFM metals Fe$_5$GeTe$_2$ [72] and Fe$_3$GaTe$_2$ [73], which can be modified by appropriate doping. Fe$_n$GeTe$_2$ and Fe$_n$GaTe$_2$ ($n$ = 3, 4, 5) have already been employed as FM electrodes in conventional MTJs [73-80]. Implementing their AFM phases in prototype-III AFMTJs is thus straightforward using similar techniques.

Furthermore, electrodes for prototype-III AFMTJs can be fabricated from 3D AFM metals that exhibit roughness-insensitive surface (interface) magnetization due to symmetry [65-67]. In these antiferromagnets, the surface magnetization is coupled to the Néel vector, so switching the Néel vector reverses the interface magnetic moments, resulting in the TMR effect predicted in this work. This significantly broadens the range of material suitable for prototype-III AFMTJs.

Importantly, from a device application perspective, writing the appropriate magnetic state in prototype-III AFMTJs is much more convenient than in the other two prototypes. Since the TMR effect is only weakly dependent on the AFM spacer thickness, one can design AFMTJs with an odd number of AFM monolayers in one electrode to induce finite magnetization in the AFM spacer layer. This property allows efficient switching of the Néel vector in this layer by applying a magnetic field (Fig. 5(A)).

Moreover, the write-in operation in prototype-III AFMTJs can be achieved electrically using spin torques. Deterministic switching of the Néel vector in an antiferromagnet can be produced by the Néel spin torques arising from staggered spin polarization on two magnetic sublattices [58] or by torque generated from spin polarization acting solely on a single magnetic sublattice [59]. Both approaches can be realized in vdW AFMTJs. Firstly, an out-of-plane tunneling spin current exerts a spin transfer torque expected to be primarily accumulated within the interfacial monolayer of the A-type AFM spacer, due to the large interlayer distance. This is supported by our calculations shown in Fig. S8(A) (see Supplemental Note S6) and mimics the torque on a single magnetic sublattice when the A-type AFM spacer is ultrathin, consisting of only two monolayers (Fig. 5(B)). Secondly, an in-plane current in the A-type electrode can generate Néel torques (Fig. 5(C)) via staggered Néel spin current [58] and/or the Edelstein effect [81,82]. The former is supported by the A-type stacking, while the latter arises from the $\hat{P}\hat{T}$ symmetry usually preserved in A-type antiferromagnets, as reflected in the layer-resolved spin textures shown in Fig. S8(B) (see Supplemental Note S6). Furthermore, it is noted that switching of an A-type vdW antiferromagnet via the spin Hall effect has recently been experimentally demonstrated [83].

The convenient control of the Néel vector state makes prototype-III AFMTJs promising for high-performance spintronic devices. In particular, we envision an AFM racetrack memory, where AFM domains can be nucleated and moved via the Néel spin torques (Fig. 5(D)) and detected via the TMR response from an AFMTJ integrated into the racetrack (Fig. 5(E)).

Finally, we highlight several challenges worth investigating for the practical applications of prototype-III AFMTJs. First, our calculations fixed the in-plane lattice constant of the AFMTJs to that of $Fe_4GeTe_2$, resulting in strains of -6.7% on BN and 1.5% on $PtTe_2$. While this did not affect our key results, since $PtTe_2$ remained metallic and BN retained a large band gap, the strain from lattice mismatch could produce wrinkling and buckling of the layers [84], potentially impacting device performance. Second, defects and disorder are unavoidable in real materials. Although we showed that TMR is robust to disorder by introducing effective energy level broadening within the Green's function formalism, this approach does not capture resonant tunneling effects due to localized defect states in the barrier [85-87]. Exploring these effects would be valuable in future studies. Lastly, our calculations for $PtTe_2/[Fe_4GeTe_2]_2/BN/[Fe_4GeTe_2]_2/PtTe_2$ AFMTJ reveal a TMR decrease from −300% to −40% when a bias voltage of 0.1 V bias voltage is applied. While this does not preclude practical observation of TMR, there is room for improvement to match the performance of commercial MTJs [88].

Overall, AFMTJs with spin-degenerate A-type AFM electrodes can produce a sizable TMR due to the spin-polarized tunneling current driven by the uncompensated interfacial magnetic moments coupled to the Néel vector. The A-type AFM stacking ensures the convenient switching of the Néel vector providing means to achieve parallel and antiparallel states in these AFMTJs. With these functionalities, this prototype of AFMTJs allows employing a broad range of available vdW AFM metals in high-performance spintronic devices.

**Methods**

First-principles calculations of the electronic structure are performed within the density functional theory (DFT) [89] using the projector augmented-wave (PAW) [90] method implemented in the VASP code [91,92]. The plane-wave cut-off energy is 500 eV and the k-point mesh about $12 \times 12 \times 1$ are used in the calculations. The exchange and correlation effects are treated within the generalized gradient approximation (GGA) developed by Perdew-Burke-Ernzerhof (PBE) [93]. The semiempirical DFT-D3 method parameterizing the van der Waals correction is used in calculations [94].

The non-equilibrium Green's function formalism (DFT+NEGF approach) [95,96], is used to calculate quantum transport, which is implemented in QuntumATK [97]. The cut-off energy of 80 Ry, the nonrelativistic Fritz-Haber Institute (FHI) pseudopotentials, and $\vec{k}$-point meshes of $11 \times 11 \times 101$ for $Fe_4GeTe_2$-based AFMTJs are used for self-consistent calculations. The $\vec{k}$-resolved transmission is calculated using $401 \times 401$ k-points in the two-dimensional (2D) Brillouin zone (BZ).

**Acknowledgments.** This work was supported by the National Key R&D Program of China (Grant No. 2024YFB3614101), the National Natural Science Foundation of China (Grants Nos. 12274411, 12241405, and 52250418), the Basic Research Program of the Chinese Academy of Sciences Based on Major Scientific Infrastructures (Grant No. JZHKYPT-2021-08), and the CAS Project for Young Scientists in Basic Research (Grant No. YSBR-084). E.Y.T acknowledges support from the Division of Materials Research of the National Science Foundation (NSF grant No. DMR-2316665). Computations were performed at Hefei Advanced Computing Center.